\documentclass[12pt]{iopart}
\pdfoutput=1
\usepackage[square,sort&compress,numbers]{natbib}
\usepackage{graphicx}

\begin{document}
    \title[How to probe universal features of absorbing phase transitions]{How to experimentally probe universal features of absorbing phase transitions using steady state}
    \author{Keiichi Tamai and Masaki Sano}
    \address{Department of Physics, The University of Tokyo, 7-3-1 Hongo, Bunkyo-ku, Tokyo, Japan}
    \ead{\mailto{tamai@daisy.phys.s.u-tokyo.ac.jp}}
    \begin{abstract}
        We propose experimentally feasible ways to probe universal features of absorbing phase transitions from two different approaches, both based on numerical validations. On one hand, we numerically study a probability distribution of duration/length of intervals of local inactive state in quasi-steady state, which has been very commonly used in experiments, in a case of the contact process. We show that the distributions obey the universal scaling ansatz expected from phenomenological scaling argument, but that care must be taken in order to suppress a bias caused by censoring due to a finite observation window. To demonstrate the latter point, we compare the distributions for the temporal intervals estimated through conventional histograms with those through the estimator which properly takes account of censoring and sampling bias. On the other hand, we also propose that, if a system is subject to uniform advection as is often the case with flowing systems, a correlation length and a correlation time near the transition point can be easily quantified by supplying the system with an active boundary condition. In order to support our proposal, we introduce a new model whose advection strength can be arbitrarily controlled. The results of numerical simulations on our model suggest that a correlation time, which is difficult to measure through the interval distributions without the aforementioned bias, can be measured through characteristic decay length of an order parameter. Crossovers between two different power-law behaviors are also identified in this case, and the universal scaling ans\"{a}tze associated with the crossovers are discussed.
    \end{abstract}
    
    
    \submitto{\JSTAT}
    \maketitle
    \section{Introduction}\label{introduction}
    
    Phase transitions to absorbing state, which the systems may enter but cannot escape from, has attracted considerable interest in literature \cite{HenkelBook, Hinrichsen2000}. This is partly because they are believed to be ubiquitous in Nature; one can find an analogy to various types of natural phenomena, including synchronization of locally coupled oscillators \cite{Baroni2001, Ahlers2002}, spatiotemporal intermittency \cite{Pomeau1986} and spreading of epidemics and forest fires \cite{Harris1974}. Another important reason is that the notion of universality, which was proven to be useful for understanding second-order phase transitions in equilibrium, has turned out to be also applicable to this type of phase transitions, despite the fact that violation of the detailed balance renders the transitions intrinsically non-equilibrium. Thanks to the universality, absorbing phase transitions are characterized by a well-defined set of universal critical exponents which does not depend on details of the system, allowing us to classify a wide variety of absorbing phase transitions into a small number of universality classes \cite{Odor2004}.
    
    One of the most well-studied universality classes of absorbing phase transitions is that of the directed percolation (DP) \cite{HenkelBook, Hinrichsen2000}. Since DP was proposed by Broadbent and Hammersley in 1957 \cite{Broadbent1957} as a model of fluid percolating through porous media, the same universal features were found in stochastic models in various contexts, such as epidemic spreading \cite{Harris1974}, catalytic reactions \cite{Ziff1986} and phenomenology of hadronic interactions at ultra-relativistic energies \cite{Cardy1980}, to name only a few. This led Janssen and Grassberger to conjecture that a system which exhibits a transition to a unique absorbing state characterized by scalar order parameter falls into the DP universality class unless the system possesses extra symmetries or conservation laws \cite{Janssen1981, Grassberger1982}. Later this robustness has been substantiated by field-theoretic renormalization group theory \cite{Janssen2001}, and numerical evidence \cite{Jensen1990, Park1993,  Zhuo1993, Munoz1996, Hoyuelos1997} indicate that the DP universality class may be even more robust than it had been conjectured to be. Thus the DP universality class is extremely robust in theory, although the critical phenomena of DP are also known to be sensitive to certain kind of perturbations such as quenched disorder \cite{Jensen1996, Janssen1997}. 

    Despite the theoretical robustness outlined above, considerably large deviations from DP were reported in many experimental studies where one would na\"{i}vely expect DP-class transition \cite{Daviaud1990, Degen1996, Rupp2003, Leipiller2007}, and experimental realization of the DP universality class had been absent for a very long time. From this, one might argue that perturbation which breaks the DP universality (such as quenched disorder) is almost always present in reality so that one can expect DP-class transitions in experiments only when some very special conditions are met. In 2007, the first convincing experimental realization of the spatially two-dimensional DP universality class has been reported using turbulent liquid crystals \cite{Takeuchi2007, Takeuchi2009}. This finding shows that the DP universality is realizable. Still, turbulent liquid crystals are substantially different from systems studied before 2007 in several ways. In particular, the absorbing state in the turbulent liquid crystals (namely, dynamic scattering mode 1 (DSM1)) is a fluctuating turbulent state, which may kill long-range interactions. Hence it remains unclear at this stage whether the DP universality class can be realized in a wide variety of experimental systems or (some of) special features of turbulent liquid crystals are crucial for observing the DP scaling.
        
    More recently, experimental evidence suggesting that the transitions to sustained turbulence in shear flow may fall into the DP universality class has been provided \cite{Sano2016, Lemoult2016}. From a viewpoint of statistical physics, one can learn two important points from these findings: First, experimental realizations of the DP universality class in two distinct spatial dimensions have been carried out. The spatially one-dimensional DP universality class was realized in the narrow Taylor--Couette flow \cite{Lemoult2016}, whereas the spatially two-dimensional one was reported in turbulent liquid crystals \cite{Takeuchi2007, Takeuchi2009} and channel flow \cite{Sano2016}. Moreover, an absorbing state does not need to be turbulent to obtain the DP scaling. These implications suggest that the DP universality class can be experimentally realized in a wider variety of systems than it had been considered to be. 
    
    Then, why was the DP universality class hard to realize in experiments? Here we would like to address this question from a methodological point of view. In numerical simulations of stochastic models, a dynamic protocol in which one tracks the dynamics starting from a certain initial condition is frequently used, two of the most widely-used initial conditions being a single active patch (in the critical spreading protocol) and a fully active state (in the critical quenching protocol) \cite{Takeuchi2014}. This is reasonable given that time is involved as an independent degree of freedom in non-equilibrium statistical physics. When applying these approaches to laboratory experiments, however, one typically suffers from lack of statistics and from slow response to the change of the control parameter (a set of experiments using turbulent liquid crystals \cite{Takeuchi2009} is a notable exception as of this writing). Hence one usually relies on a static protocol in this case: One performs a set of measurements on a quasi-steady state where physical quantities such as an order parameter fluctuate around some stationary value. In this protocol, probability distribution of the lengths (durations) of intervals of local inactive state in a quasi-steady state (hereafter referred to as ``spatial (temporal) interval distribution'') is widely used \cite{Daviaud1990,Degen1996,Rupp2003,Leipiller2007} to measure the correlation length (time) and thereby the associated critical exponent $\nu_{\perp}$ ($\nu_{||}$). Unfortunately, one generally underestimates the correlation length (time) using the interval distributions unless one carefully takes account of a fact that measurements are subject to censoring due to a finite observation window, as we will demonstrate below. Hence (apparent) deviations from the theory reported in earlier experimental works might be due to the underestimation, rather than physical characteristics of the systems. In either case, methods which allow one to correctly measure the correlation length (time) from a quasi-steady state are needed to investigate how far the notion of the universality in absorbing phase transitions can be successfully applied in reality, even though one has already encountered a few experimental realizations of the DP universality class.

    In this work, we propose practical solutions to this issue from two different approaches, both based on numerical validations. On one hand, we numerically study spatial (temporal) interval distribution of the contact process and we overcome the difficulty of underestimation by successfully employing the non-parametric estimator for bivariate recurrence time distribution \cite{Wang1999, Huang2005}, which was originally proposed in a context of medical statistics. As a result, we numerically show that the interval distributions estimated in a suitable manner satisfy universal scaling ansatz expected from phenomenological scaling theory \cite{HenkelBook} at least to a good approximation.

        The second approach, on the other hand, is to consider a different experimental procedure: We propose that applying uniform advection and active boundary condition helps one to reliably measure a set of static critical exponents via steady-state measurement, at least in a case of the DP universality class. This proposal is also motivated by the fact that advection is inevitably accompanied to spatiotemporal dynamics in some experimental systems where testifying the critical phenomena of DP is an issue of considerable interest (e.g. transitions to turbulence in pipe flow and channel flow): In such cases, measurements in moving coordinate are cumbersome, but meanwhile it is not clear from the literature how one can correctly characterize the critical phenomena of the system through measurements at fixed positions. In order to clarify these points by a numerical simulation, we introduce a new model whose advection strength can be arbitrarily controlled. The extensive study on our model reveals that a power-law scaling of a correlation length and a correlation time can be respectively examined through a characteristic decay time of a temporal interval distribution and a characteristic decay length of an order parameter. It means that one can effectively sidestep the difficulty of estimating a correlation time through interval distributions. The universal scaling ans\"{a}tze, we believe, are useful when one attempts to experimentally examine the universal features beyond a na\"{i}ve comparison of the estimated critical exponent to the theoretical value.
    
    Since the former approach is concerned with a statistical analysis of data and the latter with an actual experimental procedure, the two approaches presented in this work can be regarded as complementary. We believe this work paves a way for quantitatively studying absorbing phase transitions in reality (be it in or out of a laboratory) beyond a heuristic comparison to a toy model.
    
    The rest of this paper is organized as follows: First, we elucidate the method to measure critical exponents in an ordinary setup by measuring interval distribution in the contact process in section \ref{ordinary-case}. Particular attention is paid to appropriately estimate the interval distribution from data obtained under a finite window. After having discussed utility and weakness of steady-state measurement in an ordinary setup, we introduce the DP model with active boundary and advection in section \ref{advecsetup}. The distance dependence of the order parameter is studied in section \ref{mean-field-op}. The temporal interval distribution in the model is studied in section \ref{interval-theory}. Finally, we give concluding remarks in section \ref{discussion}.

    \section{Measurement of critical exponents in an ordinary setup}\label{ordinary-case}
    
\begin{figure}[tbp]
        \centering
        \includegraphics[width=\textwidth]{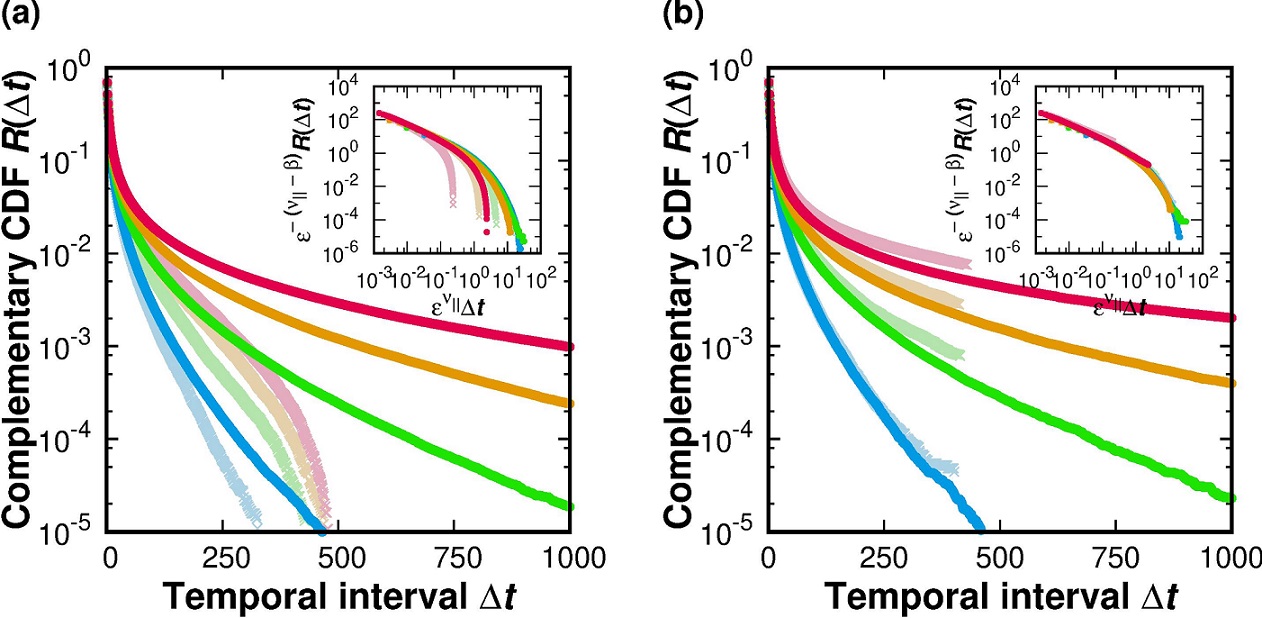}
        \caption{Temporal interval distribution of spatially one-dimensional contact process for $\lambda=3.33785$ (magenta), $\lambda=3.41098$ (orange), $\lambda=3.52412$ (green), and $\lambda=3.75039$ (blue). The distribution was estimated via (a) the empirical estimator ((\ref{empirical-estimator}) in the main text) and (b) the estimator by Wang and Chang \cite{Wang1999}. In both cases, results of the estimation under the window of $W=500$ is painted in light color whereas ones under the window of $W=5,000$ is painted in vivid color, and the inset shows the result of rescaling of the plot according to the scaling ansatz (\ref{interval-scalinghypo-CDF}). The theoretical values of the critical exponents in the DP universality class were used to achieve the data collapse.}
        \label{cptempscaling}
    \end{figure}

\begin{figure}[tbp]
        \centering
        \includegraphics[width=\textwidth]{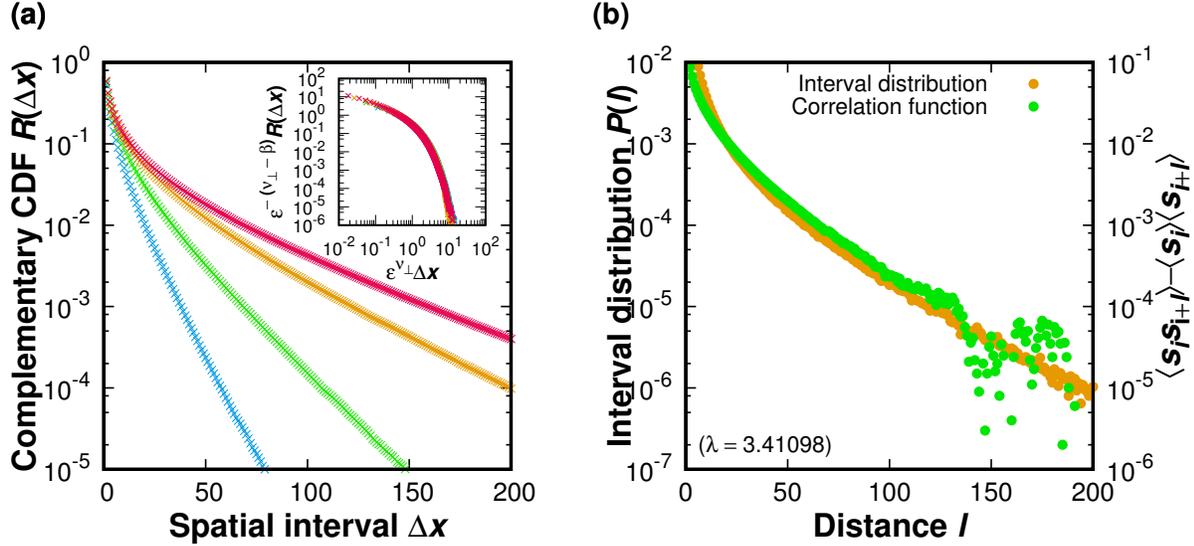}
        \caption{Spatial interval distribution of the spatially one-dimensional contact process. (a) Spatial interval distribution of spatially one-dimensional contact process for $\lambda=3.37785$ (magenta), $\lambda=3.41098$ (orange), $\lambda=3.52412$ (green), and $\lambda=3.75039$ (blue). The distribution was estimated via the empirical estimator with the observation window of $W=1,000$, and it is not problematic because the correlation length is much shorter than the window. The inset shows the result of rescaling of the plot according to the scaling ansatz (\ref{interval-scalinghypo_spatial-CDF}). The theoretical values of the critical exponents in the DP universality class were used to achieve the data collapse. (b) Spatial interval and static correlation function at $\lambda=3.41098$ are compared.}
        \label{cpspatscaling}
    \end{figure}

    We shall begin this section with brief overview of some expected properties of interval distributions. As we mentioned in section \ref{introduction}, probability distribution of the lengths (or durations) of intervals of local inactive state has been quantity of central interest in earlier experimental studies. Extensive use of the distribution is based on a belief that critical behavior of absorbing phase transitions is characterized by only two diverging length scales, namely a correlation length and a correlation time. Since local inactive state does not spontaneously turn active and the state which entire system is inactive is absorbing, the characteristic length (resp.\ time) of the decay of the distribution diverges at the critical point. Assuming that the belief is indeed the case, it is natural to expect that the characteristic length (resp.\ time) scale of the distribution corresponds to the correlation length (resp.\ time). Furthermore, if we postulate that the distribution has scaling properties analogous to those of the ordinary two-point correlation function, phenomenological scaling argument leads us to the following scaling \textit{ansatz} for temporal interval distribution $P_{||}(\Delta t;\varepsilon)$ and the spatial one $P_{\perp}(\Delta x;\varepsilon)$ \cite{HenkelBook}:
    \begin{equation}
    \label{interval-scalinghypo}
    P_{||}(\Delta t;\varepsilon)\sim\Delta t^{-(2-\beta/\nu_{||})}f_{||}(\Delta t^{1/\nu_{||}}\varepsilon),
    \end{equation}
    \begin{equation}
    \label{interval-scalinghypo_spatial}
    P_{\perp}(\Delta x;\varepsilon)\sim\Delta x^{-(2-\beta/\nu_{\perp})}f_{\perp}(\Delta x^{1/\nu_{\perp}}\varepsilon);
    \end{equation}
    where $\varepsilon$ denotes the deviation from the criticality and $f_{||},f_{\perp}$ denotes the universal scaling function for the temporal (resp.\ spatial) interval distributions. Although the interval distributions should not be regarded as an ordinary two-point correlation functions but as multi-point ones \cite{HenkelBook} in principle, one can relate the interval distributions with the two-point correlation function in a framework of a cluster mean-field-type approximation \cite{Kim2015}, where the length of the adjacent intervals are assumed to be independent from each other. Using this approach, one arrives at the same scaling ans\"{a}tze (\ref{interval-scalinghypo}) and (\ref{interval-scalinghypo_spatial}). If one is interested in considering complementary cumulative distribution function (CDF) $R(\Delta t;\varepsilon)$ defined by
    \begin{equation}
    R(\Delta t;\varepsilon):=\int_{\Delta t}^{\infty}\mathrm{d}\Delta t^{\prime} P(\Delta t^{\prime};\varepsilon),
    \end{equation}
    it follows from (\ref{cptempscaling}) and (\ref{cpspatscaling}) that
    \begin{equation}
    \label{interval-scalinghypo-CDF}
    R_{||}(\Delta t;\varepsilon)\sim \varepsilon^{\nu_{||}-\beta}h_{||}(\varepsilon^{\nu_{||}}\Delta t),
    \end{equation}
    \begin{equation}
    \label{interval-scalinghypo_spatial-CDF}
    R_{\perp}(\Delta x;\varepsilon)\sim \varepsilon^{\nu_{\perp}-\beta}h_{\perp}(\varepsilon^{\nu_{\perp}}\Delta x).
    \end{equation}
    Since the derivation of the ans\"{a}tze (\ref{interval-scalinghypo}), (\ref{interval-scalinghypo_spatial}) relies on the phenomenological scaling argument or mean-field-type approximation, it is desirable to check the validity of the ans\"{a}tze by numerical simulation. However, there has been no published work which addressed this issue to our knowledge, besides the power-law decay of the spatial interval distribution with the exponent $2-\beta/\nu_{\perp}$ expected in the vicinity of the critical point \cite{Dickman2005,Kim2015}. In the following, we mainly focus on examining the scaling ans\"{a}tze (\ref{interval-scalinghypo}) and (\ref{interval-scalinghypo_spatial}) (or equivalently, (\ref{interval-scalinghypo-CDF}) and (\ref{interval-scalinghypo_spatial-CDF})) in a case of the DP universality class.
    
    We performed Monte--Carlo simulation on the contact process in a spatially one-dimensional ring (lattice system with periodic boundary condition) with a length of 1,024 sites. The contact process \cite{Harris1974} is a continuous-time Markov process, in which each site $i$ takes either of the two states, namely an active state ($s_i(t)=1$) or an inactive state ($s_i(t)=0$). In each site, transitions from an active state to inactive one occur at rate of unity, while the reverse transitions occur at rate of $\lambda n/2d$, where $\lambda$ is the control parameter, $d$ is the number of spatial dimensions of the system, and $n$ is the number of active sites within its nearest neighbor. In a case of $d=1$, the critical point has been very precisely estimated to be $\lambda_c=3.297848(20)$ \cite{Jensen1992}. We varied the control parameter $\lambda$ in a range of $10^{-3}\lambda_c<\lambda-\lambda_c<3\times 10^{-1}\lambda_c$ and we performed $m=50$ realizations for each $\lambda$. In each realization, the system was initially set to the state where all sites are active, and we evolved the system according to the standard procedure \cite{MarroBook} for $10^7$ steps (first $10^6$ steps were discarded in a course of analysis to ensure that the quasi-steady state is measured): At each time step, we randomly selected one active site, and the selected site was deactivated with probability $1/(1+\lambda)$ while one of its nearest neighbors was selected and activated (if inactive) with complementary probability $\lambda/(1+\lambda)$, and then the time was updated as $t\rightarrow t+1/N_{\mathrm{act}}$ where $N_{\mathrm{act}}$ is the number of active sites before the attempt was given.
    
    In order to acquire an ensemble of configurations, we adopted the method by Dickman and Martins de Oliveira \cite{Dickman2005}: We first saved the $M_s=2000$ samples of configuration for each time step during the first $M_s$ steps, and then the list was updated with probability $p_{srep}=0.005$ by replacing a randomly chosen configuration on the list by the current one whenever the time step was increased by unity. In this way, 100,000 configurations were recorded. Time series of a certain site with length of $W$ time steps was also accumulated in a similar manner: Time series of randomly chosen $M_t=500$ sites were saved during the first $W$ steps, and the list was updated with probability $p_{trep}=0.05$ by replacing a randomly chosen time series on the list by the most recent time series obtained at a randomly chosen site for each $W$ time steps. Thus we collected 25,000 time series for each $\lambda$. These samples of configuration and time series were analyzed to obtain the interval distributions. 
    
    A central challenge here is to estimate the interval distribution from experimental (or numerical) data. When the observation is performed under finite observation window, as is always the case with experiments, the last interval observed within a window is never complete: Since the end of the last interval is outside the observation window, there is no way to figure out the exact length of the last interval from the data (hereafter we will refer to this situation as ``the last interval is censored''). In the presence of such a censoring, dependency between adjacent intervals is ``induced'' in a sense that the longer the first interval is the more likely the second interval is censored by the edge of the observation window (and so on) \cite{Zhu2014}, and this causes biased results if one na\"{i}vely constructs a histogram to estimate the interval distributions. To see this, let us focus on temporal interval distributions for a while. In earlier literature, the authors used to simply count the number $N_{||}(\Delta t;\varepsilon)$ of completely observed temporal intervals of length $\Delta t$. Obviously $N_{||}$ is proportional to the following empirical estimator $\hat{P}_{||}(\Delta t;\varepsilon)$ of $P_{||}(\Delta t;\varepsilon)$:
    \begin{equation}
        \label{empirical-estimator}
        N_{||}(\Delta t;\varepsilon)\propto\frac{N_{||}(\Delta t;\varepsilon)}{\sum_{\Delta t}N_{||}(\Delta t;\varepsilon)}=:\hat{P}_{||}(\Delta t;\varepsilon).
    \end{equation}
    Since this empirical estimator $\hat{P}_{||}$ is easy to construct and it converges to the intrinsic interval distribution $P_{||}(\Delta t;\varepsilon)$ of the system in a limit of $W\rightarrow\infty$ by construction, one may be tempted to rely on it. However, our results on a complementary CDF
    \begin{equation} 
        \hat{R}(\Delta t;\varepsilon):=\sum_{\Delta t^{\prime}=\Delta t}^{W}\hat{P}_{||}(\Delta t^{\prime};\varepsilon)
    \end{equation}
    for $W=500$ and $W=5,000$ reveal that the empirical estimator is very sensitive to the choice of $W$ especially near the critical point (figure \ref{cptempscaling}(a)). This implies that the empirical estimator is an artificial quantity in a sense that it strongly depends on a parameter which can be determined at observers' will regardless of the dynamics of the system, and one can no longer expect the scaling ans\"{a}tze (\ref{interval-scalinghypo}) or (\ref{interval-scalinghypo_spatial}) to hold. Indeed, the rescaled complementary CDFs for various $\lambda$ do not overlap in general as shown in the inset of figure \ref{cptempscaling}(a), although the scaling collapse is observed at far away from the critical point for $W=5,000$. Thus, relying on the empirical estimator can easily lead one to biased results. As a rule of thumb, an observation window needs to be, at the shortest, 10 times longer than the correlation time of the system so that the empirical estimator works properly. What is worse, taking an extremely large observation window is usually infeasible due to difficulty in keeping the system temporally/spatially homogeneous (in a practical sense) in such a wide range, making the estimation of the distribution from data taken under limited window size an inevitable challenge. 
    
    Actually, one of the reasons why Takeuchi \textit{et al.}\ managed to demonstrate the divergence of the correlation length/time in agreement with the DP universality class even within steady-state measurements is that they introduced a correction to suppress the artifact caused by a finite observation window: They divided the number $N(\Delta l)$ of observed spatial (temporal) intervals of size $\Delta l$ by $1-\Delta l/W$. This correction is based on a somewhat heuristic argument saying that an interval of size $l$ may not be captured within a frame of size $W$ with probability $l/W$. Unfortunately, this correction, again, does not properly take account for the induced dependency between adjacent intervals, and the resulting interval distributions are still biased, although practical applicability of the method seems much better than the empirical estimator (see also Appendix). 
    
    The above considerations motivate us to look for an alternative way to estimate interval distributions, which is insensitive against the choice of $W$ as long as the number $m$ of realizations is sufficiently large. Here, we employ a technique for estimating univariate recurrence time distribution proposed by Wang and Chang \cite{Wang1999}. The estimator for the complementary CDF by Wang and Chang is similar to a well-known Kaplan--Meier estimator \cite{Kaplan1958, AndersenBook}, but they carefully avoid a sampling bias by discarding the last (possibly censored) interval unless there is only one (censored) interval in a time series. Formally stating, the estimator $\hat{R}(t)$ is given by the following:
    \begin{equation}
    \hat{R}(t):=\prod_{t^*\le t}\left(1-\frac{D(t^*)}{M(t^*)}\right);
    \end{equation}    
    \begin{equation}
        D(t):=\sum_{i=1}^m \Theta(S_i-1)\cdot\frac{N_i(t)}{S_i},
    \end{equation}
    \begin{equation}
        M(t):=\sum_{i=1}^mM_i(t) \quad \mathrm{with}\quad M_i(t)=\left\{
        \begin{array}{cl}
        \Theta(W-t) & S_i=0\\
        \sum_{t^{\prime}=t}^{W}N_i(t^{\prime})/S_i & \mathrm{otherwise}\\
        \end{array}
        \right.
        ,
    \end{equation}
    where $\Theta$ is the standard Heaviside step function with $\Theta(0)=1$, $N_i(t)$ represents the number of completely observed intervals of length $t$ within the $i$th realization, $S_i(=\sum_{t^{\prime}}N_i(t^{\prime}))$ represents the total number of completely observed intervals within the $i$th realization, and $t^*$ is the ordered and distinct uncensored times. Nice thing about this estimator is that, assuming conditional independence and independent censoring (that is, the censoring occurs regardless of the dynamics of the system), one can theoretically show that the estimator weakly converges to the intrinsic distribution $R(t)$ of the system \cite{Wang1999}. Note also that we can still apply this technique even though time series of current interest is not univariate but consists of two states observed alternatingly \cite{Huang2005}, which is the case for the present study. We calculated the above estimator in the contact process for $W=500$ and $W=5,000$, and we found that the result is much less sensitive to the choice of the value of $W$ than the empirical estimator is (figure\ \ref{cptempscaling}(b)), although slight difference is still present. We also tested the scaling ansatz (\ref{interval-scalinghypo}) and we obtained a reasonable collapse both for $W=500$ and $W=5,000$  (inset of figure\ \ref{cptempscaling}(b)).
    
    Although the same argument also applies to spatial interval distribution in principle (and hence we do not repeat it explicitly), it is worthwhile to emphasize that the situation is much simpler in practice. Since $\nu_{\perp}<\nu_{||}$ holds in the DP universality class (as well as in other universality classes of absorbing phase transitions known so far \cite{HenkelBook}), a correlation length grows much slower than a correlation time as getting close to the critical point. Suppose, for example, one performs a set of experiments in a spatially one-dimensional system with control parameter $\varepsilon$ varied over two orders of magnitude to provide reliable estimates of the critical exponents (in accordance with a critical remark by Stumpf and Porter \cite{Stumpf2012}). Then, the correlation length of the system grows about 150 times in a course of measurements whereas the correlation time about 3,000 times(!), if the system does fall into the DP universality class. This difference of one digit drastically changes the difficulty in measuring critical exponents through interval distributions. Indeed, we confirmed in figure \ref{cpspatscaling}(a) that the expected scaling ansatz (\ref{interval-scalinghypo_spatial-CDF}) is numerically reproduced even if we perform measurements in a system with moderate size (for a numerical simulation, at least) using the empirical estimator.
    
    Provided that the issue of estimation is made less extreme, the interval distribution serves as a very useful tool to characterize collective dynamics off the critical point. The utility of interval distribution can be seen by comparing it with the ordinary static two-point correlation function $C_s(l):=\langle s_i(t)s_{i+l}(t)\rangle-\langle s_i(t)\rangle\langle s_{i+l}(t)\rangle$: The spatial interval distribution and the static spatial correlation function at $\lambda=3.41098$ (that is, $\varepsilon\sim 4\times 10^{-2}$) is shown in figure \ref{cpspatscaling}(b) as an example. The results suggest that both the interval distribution function and the correlation function decay exponentially and that the characteristic length of the decay coincides with each other. A significant difference can be found for $l\sim 130$, where the correlation function begins to scatter presumably due to lack of statistics. Thus, the severe demand of statistics makes the measurement of the correlation function impracticable in typical experiments, and the interval distribution function is an attractive alternative in that case.

    Nevertheless, some critical remarks have to be made before closing this section: When we rescaled the data according to the scaling hypotheses (\ref{interval-scalinghypo-CDF}) and (\ref{interval-scalinghypo_spatial-CDF}), we made use of prior knowledge of the critical point. In practical situations, however, one has to determine the location of the critical point separately. While the critical quenching protocol where one begins the experiment with fully active state and tracks the time evolution of the activity is a good choice for this purpose (if available), this protocol is often infeasible due to difficulty in quickly changing the control parameter of the system. If the quenching protocol is not available, one is forced to rely on the stationary value of the order parameter, but then one has to deal carefully with an accidental fall into an absorbing state. It is also important to note that validity of the estimator is not obvious \textit{a priori} because lengths adjacent intervals may not be independent of each other, although it could be justified \textit{a posteriori} just as we demonstrated in the present study. It is therefore desirable to find another experimental procedure which sets one free from the aforementioned issues even if only steady-state measurements are available.
    
    \section{Measurement of critical exponents in flowing system}\label{numerics}
    The main purpose of this section is to demonstrate that introducing an active boundary condition and uniform advection helps one to efficiently measure critical exponents via steady-state measurements in the case of DP universality class. As will be clarified in the following, the difficulty of measuring $\nu_{||}$ is worked around by probing the spatial decay of the order parameter (Note that, since the boundary condition breaks the spatial translation-invariance, the order parameter is no longer uniform even in the stationary state). Also, it will be demonstrated that temporal interval distribution, which was studied in the previous section in an ordinary setup, can serve as a reliable means to measure the remaining static critical exponent $\nu_{\perp}$. To see this, we introduce a new model, perform an extensive Monte--Carlo simulation, and provide a theoretical interpretation of the numerical results, either in heuristics or mean-field theory.
    
    \subsection{Description of the model}\label{advecsetup}
\begin{figure}[tbp]
        \centering
        \includegraphics[width=\textwidth]{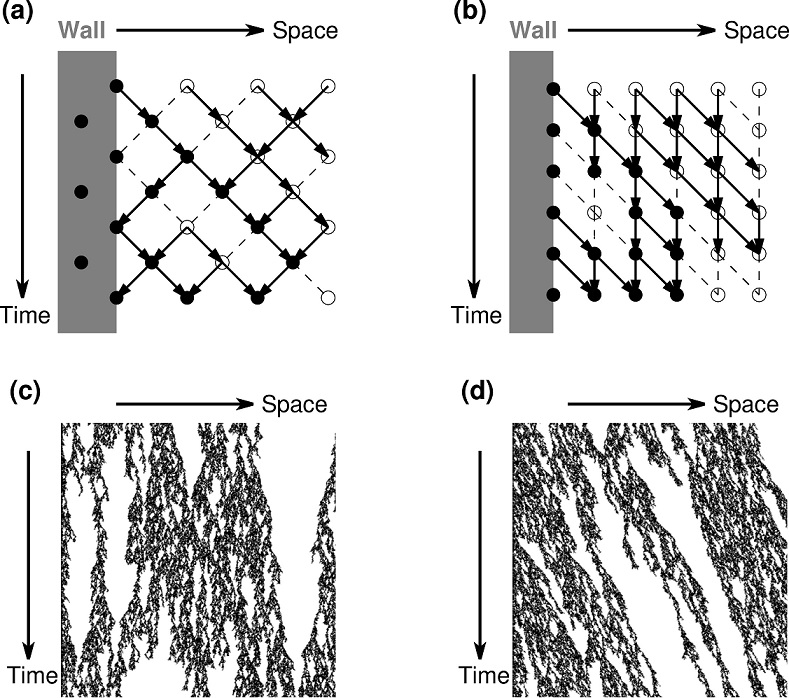}
        \caption{Scheamtic representation of the model. (a) A case of $\theta=0$, where the rules (\ref{ActiveWall-advecrule}) and (\ref{ActiveWall-noadvecrule}) are alternatingly applied at each time step (see the main text). (b) A case of $\theta=\pi/4$, where the rule (\ref{ActiveWall-advecrule}) is applied at every time step. For (a) and (b), open bonds and closed ones are respectively represented in solid arrows and dashed lines. The leftmost site is forced to be active (filled circle) by the boundary condition, and sites connected with at least one active neighbouring site by an open bond are defined to be active. Inactive sites are represented in empty circles. (c) Typical spatiotemporal dynamics near the wall for $\theta=0$. (d) Typical spatiotemporal dynamics near the wall for $\theta=2\pi/9$. For (c) and (d), a case with $p=0.64807$ ($\varepsilon:=(p-p_c)/p_c\sim 5\times 10^{-3}$) is shown.}
        \label{slanted-wall_schematics}
    \end{figure}

    Our model consists of a string $\{s_i\}_{i=0}^{L-1}$ of $L$ sites each of which can take either of two states, namely an active state ($s_i=1$) or an inactive state ($s_i=0$). In order to mimic the effect of advection, we introduce a new parameter $\theta$: If the integer part of $t\tan\theta$, which indicates how far one is carried by the advection, changes after the increment of $t$ by 1 (that is, $\lfloor(t+1)\tan\theta\rfloor=\lfloor t\tan\theta\rfloor+1$, where $\lfloor x \rfloor$ is the standard floor function), we apply an advection by introducing the following transition rules when $t$ is an integer:
    \begin{equation}
        \label{ActiveWall-advecrule}
        \begin{array}{c}
        P(s_i(t+1/2)=1|s_i(t)=1,s_{i-1}(t)=1)=1-(1-p)^2,\\
        P(s_i(t+1/2)=1|s_i(t)=1,s_{i-1}(t)=0)=p,\\
        P(s_i(t+1/2)=1|s_i(t)=0,s_{i-1}(t)=1)=p,\\
        P(s_i(t+1/2)=1|s_i(t)=s_{i-1}(t)=0)=0\\    
        \end{array}
    \end{equation}
    instead of the standard rule of the directed bond percolation for an integer $t$:
    \begin{equation}
        \label{ActiveWall-noadvecrule}
        \begin{array}{c}
        P(s_i(t+1/2)=1|s_i(t)=1,s_{i+1}(t)=1)=1-(1-p)^2,\\
        P(s_i(t+1/2)=1|s_i(t)=1,s_{i+1}(t)=0)=p,\\
        P(s_i(t+1/2)=1|s_i(t)=0,s_{i+1}(t)=1)=p,\\
        P(s_i(t+1/2)=1|s_i(t)=s_{i+1}(t)=0)=0.\\    
        \end{array}
    \end{equation}
    Time $t$ is increased by $1/2$ in each step after all the sites are updated, and we always evolve the system according to the rule (\ref{ActiveWall-advecrule}) when $t$ is a half-integer. Schematic representation of the rule for $\theta=0$ and $\theta=\pi/4$ can be respectively found in figure \ref{slanted-wall_schematics}(a) and figure \ref{slanted-wall_schematics}(b). In the present work, we restrict $\theta$ not to be larger than $\pi/4$.

    When one considers an impact of advection, a boundary condition plays a crucial role. If we impose a periodic boundary condition as one usually does in numerical simulation, advection can be transformed away by Galilean transform and therefore advection is not expected to affect essential features of the dynamics of the model. Here we impose an active boundary condition at the wall instead: 
    \begin{equation}
    \label{numerical-boundary}
    s_0(t)=1\ \mathrm{for}\ \forall t.
    \end{equation}
    This is one of the simplest setups in which advection may affect the dynamics in a non-trivial manner. Another important modification the active boundary condition provides is that the system no longer possesses unique absorbing state, and hence that a non-trivial steady state is well-defined even for finite system \cite{Howard2000, BlythePhD}. This is in a sharp contrast with the ordinary setup we discussed in a previous section, where, for a finite system, the true steady state is always absorbing and one is often forced to consider the \textit{quasi}-steady state. When necessary, an absorbing boundary condition was imposed on the other side of the system.
    
    We performed a Monte--Carlo simulation on a lattice of the size $N$. Unless otherwise stated, we set $N=8,192$. We started our simulation with the system whose site in the wall is active and other sites are inactive. In order to study the steady state of the model, we first performed the simulation for $W_{\mathrm{warm}}$ timesteps and then the statistics were accumulated over another $W_{\mathrm{sample}}$ timesteps. Unless otherwise stated, we fixed $W_{\mathrm{warm}}$ and $W_{\mathrm{sample}}$ to be $W_{\mathrm{warm}}=15\times 10^4$ and $W_{\mathrm{sample}}=85\times 10^4$, respectively. In most cases, the warm-up time $W_{\mathrm{warm}}$ was long enough for the wave of the activity (similar to what was reported in Ref.\ \cite{Costa2010}) to reach to the other side of the system. The simulation was repeated over 16 times to further improve statistics.
    
    \subsection{Distance dependence of order parameter}\label{mean-field-op}
\begin{figure}[tbp]
        \centering
        \includegraphics[width=\textwidth]{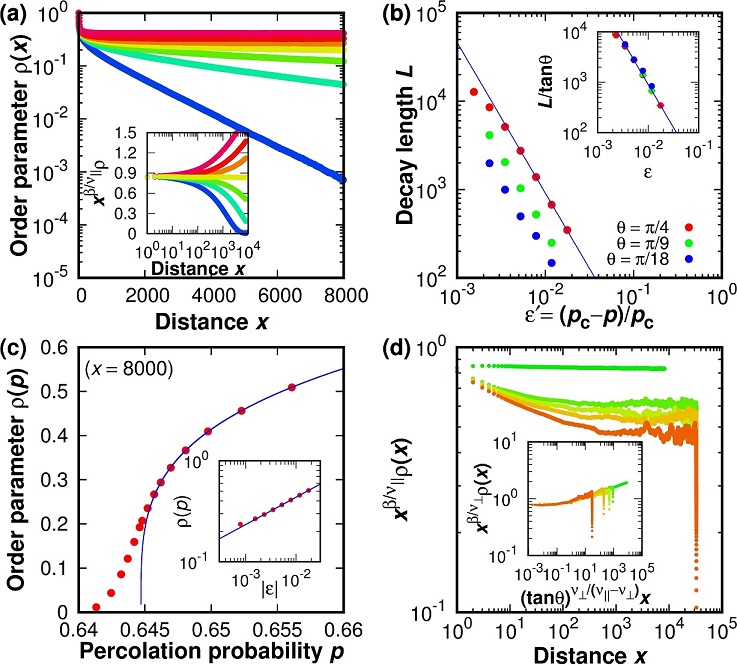}
        \caption{Order parameter of the directed percolation model with active wall and asymmetric connection. (a) Spatial dependence of the order parameter $\rho(x)$ for $p-p_c=(1.5)^{2k}\times 10^{-3}$ ($k=2,1,0$), $p=p_c$ and $p-p_c=-(1.5)^{2k}\times 10^{-3}$ ($k=0,1,2$), from top to bottom. A case with $\theta=\pi/4$ is shown as an example. Inset shows the same data, but the vertical axis is rescaled by $x^{\beta/\nu_{||}}$. (b) Decay length $L$ (see text) of the order parameter for various $\theta$. The inset shows the decay length $L$ divided by $\tan\theta$. (c) Order parameter $\rho$ measured at $x=8000$ for various $\theta=\pi/4$. Solid curve represents a guide to eye for power-law behavior with exponent $\beta$. (d) $\rho(x)$ in a vicinity of the critical point ($p=0.644710$) for $\theta=\pi/4$ and $\theta=(7-2k)\pi/180$ ($k=0,1,2,3$), from top to bottom. Here, we performed a longer numerical simulation in both space and time: $N=32768$, $W_{\mathrm{warm}}=10^6$ and $W_{\mathrm{sample}}=3\times 10^6$, and statistics were acquired over 8 realizations. Inset shows the same data, but both the horizontal axis and the vertical axis are rescaled as $(\tan\theta)^{\frac{\nu_{\perp}}{\nu_{||}-\nu_{\perp}}}x$ and $x^{\beta/\nu_{\perp}}\rho(x)$, respectively. The theoretical values of the critical exponents in the DP universality class were used to achieve the data collapse.}
        \label{ActiveWallop1D}
    \end{figure}

    First, we shall see that the model possesses two distinct phases by studying the order parameter. Due to the breakdown of spatial translation-invariance we mentioned above, order parameter $\rho$ has to be measured with respect to the distance $x$ from the wall. We defined the order parameter $\rho(x)$ as a fraction of time that the $x$th site from the wall is active during the stationary simulation. The results shown in figure \ref{ActiveWallop1D}(a) shows that $\rho(x)$ decays exponentially when the percolation probability $p$ is small whereas it saturates to a finite value when $p>p_c$. This suggests that the model has a distinct transition from so-called low-density phase to high-density phase, despite the fact that absorbing state is no longer present. Although the critical point $p_c$ is expected to be identical with the original directed bond percolation, location of the critical point can be determined \textit{a posteriori} by using the power-law decay of the order parameter which is expected near the critical point: At the critical point, the order parameter is expected to decay in a power law with exponent of $\beta/\nu_{||}$ as shown in the inset of figure \ref{ActiveWallop1D}(a).
    
    The exponential decay in a low-density phase allows us to define a decay length $L$ as a characteristic length of the decay at $p<p_c$, that is,
        \begin{equation}
          \rho(x)\sim\exp(-x/L(p;\theta))\quad \mathrm{for\ sufficiently\ large}\ x.
        \end{equation}
        As shown in figure \ref{ActiveWallop1D}(b), we found that $L$ as a function of $\varepsilon^{\prime}:=(p_c-p)/p_c$ shows a power-law behavior in a vicinity of the critical point, and the exponent of the power-law behavior coincides with critical exponent $\nu_{||}$ of DP associated with divergence of the correlation time, as long as advection is sufficiently strong (that is, $\theta$ is large). These results are consistent with those reported by Costa and his coworkers \cite{Costa2010}, where a heuristic argument supporting the results is also given. Indeed, one can see, from the inset of figure \ref{ActiveWallop1D}(b), that $L(p;\theta)/\tan\theta$ takes very similar value among various $\theta$, provided that $p$ is same. This implies that one can use the order parameter instead of interval distribution to probe divergence of correlation time, allowing him/her to effectively sidestep the difficulty we described in the previous section.
    
    The price to pay for the ability to measure correlation time using the spatial decay of the order parameter $\rho(x)$ is difficulty in measuring critical exponent $\beta$ from the stationary value of $\rho(x)$ which is expected for sufficiently large $x$. Since the decay length $L$ diverges in the same manner as the correlation time, there is a considerable fraction of time in which the site is active even in a low-density phase unless one measures $\rho$ at extremely far away from the wall. We demonstrate this effect by showing the results of the measurement of the order parameter $\rho$ at the fixed observation point (the 8,000th site from the active wall) for $\theta=\pi/4$ in figure\ \ref{ActiveWallop1D}(c). One can see a significant round-off at the onset of the high-density phase. However, we are still able to plot the stationary value with respect to $\varepsilon$ by making use of the knowledge of the location of the critical point we can obtain through $\rho(x)$, and the power-law behavior with the exponent of $\beta$ can be confirmed unless the system is too close to the critical point.

    Having seen how the critical exponents $\beta$ and $\nu_{||}$ can be measured using the steady state of this model, let us now study how $\rho(x)$ experiences crossover from a case without advection (just a simple active boundary condition, as mentioned in e.g. \cite{Howard2000, Frojdh2001}) to a case with advection. To address this issue, we measured $\rho(x)$ in a vicinity of the critical point ($p=0.644710$, that is, $\varepsilon\sim 1.5\times 10^{-5}$) for various $\theta$. Since the correlation length and the correlation time of the system is expected to be very large, we performed the simulation on a setup which is larger in both space and time: We set $N$, $W_{\mathrm{warm}}$ and $W_{\mathrm{sample}}$ respectively to be $N=32768$, $W_{\mathrm{warm}}=10^6$ and $W_{\mathrm{sample}}=3\times 10^6$ in this case. As one can see from the results shown in figure \ref{ActiveWallop1D}(d), we observe a power law decay whose exponent is different from $\beta/\nu_{||}$ that we have seen previously: Actually, the exponent coincides with $\beta/\nu_{\perp}$ which is expected from a phenomenological scaling argument in a case without advection \cite{HenkelBook}. Beyond a certain lengthscale, the power-law decay with the exponent $-\beta/\nu_{||}$ is recovered, although deviation from the expected behavior is present due to an inactive boundary condition placed at the other side of the system.
    
    A mean-field theory provides a useful way to interpret the crossover. Although the mean-field theory is not exact in a $d$-dimensional system with $d\le 3$, it successfully captures the essential features of our results. We focus on a case where the spatially $d$-dimensional system is supplemented with a fixed active wall of $(d-1)$-dimensional plane and advection perpendicular to the wall (hereafter refer to this direction as ``$x$-axis''). In a framework of the mean-field theory, the system is described by the following \cite{Costa2010}:
    \begin{equation}
        \label{noiseless-DPLangevin}
        \frac{\partial \rho}{\partial t}=\varepsilon\rho-\lambda\rho^2-v_x\frac{\partial\rho}{\partial x}+D\nabla^2\rho;
    \end{equation}
    where  $\varepsilon$, $\lambda$ and $D$ are phenomenological parameters, and $v_x$ represents the velocity of the advection. Boundary condition is set to
    \begin{equation}
        \label{active-boundary-condition}
        \lim_{x\rightarrow 0}\rho(x)=\infty.
    \end{equation}
    Note that, even in the mean-field theory, the fourth term of the RHS of (\ref{noiseless-DPLangevin}) can no longer be neglected because of the breakdown of spatial translation-invariance. If one is interested in steady state, one may put $\partial\rho/\partial t=0$, and since spatial translation-invariance is expected to hold except for streamwise direction, one may also put $\nabla^2\rho=\partial^2_x\rho$. As a result, we obtain the following Fisher--Kolmogorov--Petrovskii--Piskunov (FKPP) equation:
    \begin{equation}
        \label{DPMF-steady}
        \varepsilon\rho-\lambda\rho^2-v_x\frac{\mathrm{d} \rho}{\mathrm{d} x}+D\frac{\mathrm{d}^2 \rho}{\mathrm{d} x^2}=0.
    \end{equation}    
    To focus on our purpose with the simplest case, we concentrate on a case with $\varepsilon=0$, that is, when the system is exactly on a critical point (asymptotic solutions for large $x$ in both high-density and low-density phase has already been given in Ref.\ \cite{Costa2010}). If the advection is absent ($v_x=0$), we can easily solve (\ref{DPMF-steady}) to find $\rho(x)=6Dx^{-2}/\lambda$. The exponent of $2$ coincides with the mean-field DP exponent of $\beta/\nu_{\perp}$. Meanwhile, we find $\rho(x)=v_xx^{-1}/\lambda$ if the effect of advection is so large ($v_x\gg 1$) that the diffusion (fourth) term of (\ref{DPMF-steady}) is negligible compared to the advection (third) term. The exponent of unity coincides with the mean-field DP exponent of $\beta/\nu_{||}$. 
    
    Given that which solution can be found is a matter of comparison between the advection term and the diffusion term, it is natural to ask for lengthscale $x_*$ at which they become comparable to each other. To address this question, we check the ratio of the terms which are assumed to be dominant over one assumed to be negligible \cite{BenderBook}. If we substitute $\rho(x)=6Dx^{-2}/\lambda$ (the solution of the (\ref{DPMF-steady}) for $v_x=0$) to the advection term and take a ratio over the diffusion term, we obtain
    \begin{equation}
        v_x\frac{\mathrm{d}}{\mathrm{d}x}\frac{6D}{\lambda}x^{-2}/\left(D\frac{\mathrm{d}^2}{\mathrm{d}x^2}\left(\frac{6D}{\lambda}x^{-2}\right)\right) = \frac{v_x}{2D}x.
    \end{equation}
    Thus, the assumption that the advection term is negligible compared to the diffusion term is valid for $x\rightarrow 0$. Conversely, a similar argument leads us to find that the the assumption that the diffusion term is negligible to the advection term is valid for $x\rightarrow\infty$. As a result, we have the following asymptotic solutions of the (\ref{DPMF-steady}): 
    \begin{equation}
        \label{rho-asymptotic}
        \rho(x)\sim\left\{
        \begin{array}{cl}
            \displaystyle \frac{6D}{\lambda}x^{-2}; & x\rightarrow 0,\\
            \displaystyle \frac{v_x}{\lambda}x^{-1}; & x\rightarrow \infty.\\
        \end{array}
        \right.
    \end{equation}
    The distance $x_*$ which satisfies $6Dx_*^{-2}/\beta=v_xx_*^{-1}/\beta$, that is, 
    \begin{equation}
    \label{MFcrossoverexponent}
    x_*=6D/v_x,
    \end{equation}
    gives an approximate length scale at which $\rho(x)$ experiences the crossover from $\rho(x)\sim 6Dx^{-2}/\lambda$ to $\rho(x)\sim v_xx^{-1}/\lambda$. As expected, $x_*$ diverges to infinity as $v_x\rightarrow +0$.
    
    Even though the mean-field theory cannot be directly applied when the spatial dimension $d$ is less than 4, one can give a heuristic argument to estimate the typical lengthscale of the crossover, albeit more roughly. Repeating the similar argument as above yields
    \begin{equation}
        \rho(x)\sim\left\{
        \begin{array}{cl}
            \displaystyle x^{-\delta_{\perp}}; & x\rightarrow 0,\\
            \displaystyle (x/v_x)^{-\delta_{||}}; & x\rightarrow\infty,\\        
        \end{array} \right.
    \end{equation}
    (where $\delta_{\perp}=\beta/\nu_{\perp}$ and $\delta_{||}=\beta/\nu_{||}$), and thereby 
    \begin{equation}
        \label{opcrossover-lengthscale}
        x_*\sim v_x^{-\frac{\nu_{\perp}}{\nu_{||}-\nu_{\perp}}}.
    \end{equation}
    It is worthwhile to note that the value of $\nu_{\perp}/(\nu_{||}-\nu_{\perp})$ is unity in $d\ge 4$, so that it coincides with what we have obtained in the mean-field theory (\ref{MFcrossoverexponent}). Recalling the empirical observation in figure \ref{ActiveWallop1D}(d) that the coefficient of the power-law decay with the exponent $\beta/\nu_{\perp}$ is not sensitive to the value of $\theta$, we arrive at the following scaling ansatz:
    \begin{equation}
        \label{opcrit-scaling}
        \rho(x;\theta)\sim x^{-\beta/\nu_{\perp}}f((\tan\theta)^{{\frac{\nu_{\perp}}{\nu_{||}-\nu_{\perp}}}}x),
    \end{equation}
    where $f$ is the universal scaling function. We rescaled the data in figure \ref{ActiveWallop1D}(d) according to this ansatz, and we found a reasonably good collapse as a result, as shown in the inset of figure \ref{ActiveWallop1D}(d) (besides the artifact due to the inactive boundary condition). Thus, using the crossover, one might be able to measure all the three static critical exponents from $\rho(x)$ in principle.
    
    A few comments are in order before closing this subsection: The decay length $L$ is significantly different from the curtain width $W$ introduced by Chen \textit{et al}. \cite{Chen1999}, although the intuitive meanings of these quantities are similar to each other. In Ref.\ \cite{Chen1999}, the authors defined the curtain width $W$ to be the mode of the probability distribution $P(w)$ that the \textit{farthest} active site is at $w$ sites away from the active wall (in fact the mode coincides with the mean in this case). If one approximates the dynamics of active clusters near the active wall as a collection of needles which are uncorrelated with each other beyond the spatial correlation length $\xi_{\perp}$, $P(w)$ is well approximated by
    \begin{equation}
        P(w)\sim P_{\mathrm{surv}}\left(\frac{n\xi_{\perp}}{\tan\theta}\right)\prod_{n^{\prime}>n}\left[1-P_{\mathrm{surv}}\left(\frac{n^{\prime}\xi_{\perp}}{\tan\theta}\right)\right],
    \end{equation} 
    where we defined an integer $n$ so that $(n-1)\xi_{\perp}\le w<n\xi_{\perp}$ and $P_{\mathrm{surv}}(t)$ is the survival probability of the active cluster which departed from the active wall at $t=0$. Notice that $P(w)$ is not simply proportional to $P_{\mathrm{surv}}(n\xi_{\perp}/\tan\theta)$ and the extra factor causes the logarithmic correction to the resulting curtain width
    \begin{equation}
        W\simeq A\varepsilon^{-\nu_{||}}\log\left(\frac{\varepsilon_0}{\varepsilon}\right),
    \end{equation}
    even if we assume the standard functional form of $P_{\mathrm{surv}}(t)$:
    \begin{equation}
        P_{\mathrm{surv}}(t)\propto \exp\left(-\frac{t}{\xi_{||}}\right).
    \end{equation}
    Meanwhile, the order parameter $\rho(x)$ in present work, defined as a fraction of time that the $x$th site from the wall is active, is nothing but $P_{\mathrm{surv}}(x/\tan\theta)$, at least in a framework of the independent needles approximation. Since neither $x$ nor $\tan\theta$ depends on the deviation $\varepsilon$ from the criticality, one can expect a simple scaling for the decay length $L$. This contrast between $W$ and $L$ can be considered as a warning to suitably choose a macroscopic characterizer when one tries to quantitatively characterize a system with an active wall and advection.
    
    Another comment concerns with apparent contradiction with Costa \textit{et al.}, who claimed that the exponent $\nu_{\perp}$ cannot be accessed from steady-state density profile. In the driven asymmetric contact process (DACP), where the asymmetric contact process \cite{Schonmann1986} is supplemented with a fixed active boundary condition, the velocity $v_x$ of the advection and the diffusion constant $D$ are coupled with the activation rate $r$ by the following:
    \begin{equation}
        v_x=r(1-\rho),\quad D=v_x/2.
    \end{equation} 
    This implies that the velocity $v_x$ of the advection is almost fixed near the critical point, and $v_x$ is not close to zero (recall that $r_c=3.3055(5)$). Also, assuming (\ref{opcrossover-lengthscale}) gives a good approximation of the lengthscale of the crossover, the typical lengthscale $x_*$ of the crossover decreases very rapidly with $v_x$. This makes the observation of the crossover near the criticality in DACP practically impossible. However, the activation rate and the velocity of the advection may be safely regarded as independent from each other in some cases (e.g.\ applying shear flow to electroconvection of liquid crystals), and the crossover could be observed in that case. Thus, our claim and that made by Costa \textit{et al.}\ are not incompatible with each other, but the apparent contradiction stems from the specific selection of the model.

    \subsection{Inactive interval distribution}\label{interval-theory}    
    Since the crossover of the order parameter is not clearly observed unless the advection is sufficiently weak, it is desirable to find an alternative way to measure the remaining static critical exponent $\nu_{\perp}$. One of the possible candidates is the velocity $v_{\mathrm{sup}}$ of the wave of activity in the high-density phase (hereafter referred to as ``the supercritical wave velocity''), as studied by Costa \textit{et al.} \cite{Costa2010}: It can be theoretically shown that $v_{\mathrm{sup}}$ obeys the following functional form:
    \begin{equation}
        \label{vsup-prediction}
        v_{\mathrm{sup}}=v_c+A\varepsilon^{\nu_{||}-\nu_{\perp}}.
    \end{equation}
    Unfortunately, numerical simulation of the driven asymmetric contact process shows some deviation from the theoretical prediction (\ref{vsup-prediction}), possibly due to significant corrections to the leading-order behavior \cite{Costa2010}. Another unhappy thing about the measurement of $\nu_{\perp}$ through the supercritical wave velocity $v_{\mathrm{sup}}$ is that the measurement would be plagued with an extra delicate issue of determining the offset velocity $v_c$ and with larger uncertainty due to a propagation of error even if the sub-leading corrections were sufficiently weak. Thus, one is in need of a practical means to measure $\nu_{\perp}$ other than $v_{\mathrm{sup}}$.
    
    In the following, we demonstrate that $\nu_{\perp}$ can be measured via the temporal interval distributions measured at far away from the wall. Although the temporal interval distribution generally depends on the distance $x$ from the active wall, choice of $x$ does not make a noticeable difference as long as the saturation of $\rho(x)$ is achieved. We measured the interval distribution using the empirical estimator (\ref{empirical-estimator}). The use of the empirical estimator is justified because the correlation time of the system is about $10^4$ time steps at the longest within the range of $p$ in the present study (recall the inset of figure \ref{ActiveWallop1D}(b)), and hence $W_{\mathrm{sample}}$ is about 100 times longer than the correlation time. 
    
\begin{figure}[tbp]
        \centering
        \includegraphics[width=\textwidth]{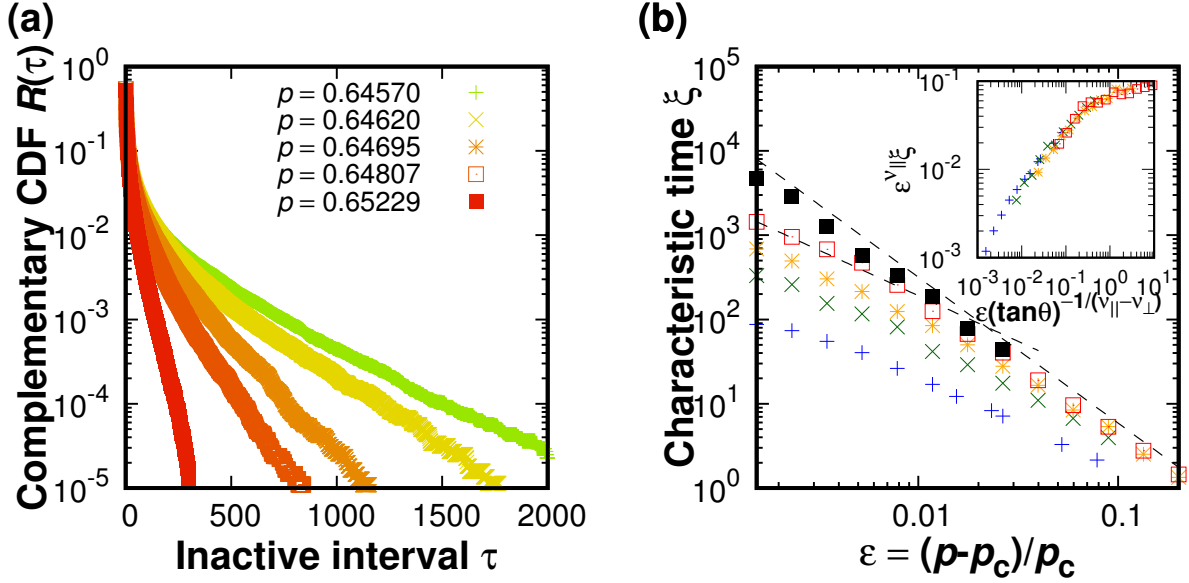}
        \caption{Temporal interval distribution of the directed percolation model with active wall and asymmetric connection. (a) Temporal interval distributions for various $p$ is shown for $\theta=\pi/12$ as an example. (b) Dependence on 
        the distance to the criticality $\varepsilon$ of characteristic time $\xi$ for $\theta=0$ (black), $\theta=\pi/36$ (red), $\theta=\pi/18$ (orange), $\theta=\pi/9$ (green) and $\theta=\pi/4$ (blue). The steeper dashed line shows a guide-to-eye for the power-law behavior with exponent $\nu_{||}$, and the other one for $\nu_{\perp}$. The inset shows the same data, but the horizontal axis and the vertical axis is rescaled as $\varepsilon(\tan\theta)^{-\frac{1}{\nu_{||}-\nu_{\perp}}}$ and $\varepsilon^{\nu_{||}}\xi$, respectively. The theoretical values of the critical exponents in the DP universality class were used to achieve the data collapse.}
        \label{decaylength}
    \end{figure}

    As with the ordinary case we have seen in section \ref{ordinary-case}, we observed a crossover to an exponential decay for sufficiently large $\tau$ (figure \ref{decaylength}(a)), enabling us to investigate a characteristic time $\xi$ of the decay:
    \begin{equation}
        R(\tau)\sim \exp(-\tau/\xi(p;\theta))\quad \mathrm{for\ sufficiently\ large}\ \tau.
    \end{equation}
    The characteristic time diverges in a power law with exponent $-\nu_{||}$ when $\theta=0$ as expected. As we increase the velocity $\tan\theta$, however, we observe that $\xi$ deviates from the expected power law, but exhibits power-law divergence with different exponent: In fact, the results shown in figure \ref{decaylength}(b) suggest that the new exponent is close to the critical exponent $\nu_{\perp}$ of the DP universality class. The larger a value of $\theta$ is, the larger a value of $\varepsilon_*$ at which $\xi$ experiences the crossover becomes.    
    
\begin{figure}[tbp]
        \centering
        \includegraphics[width=\textwidth]{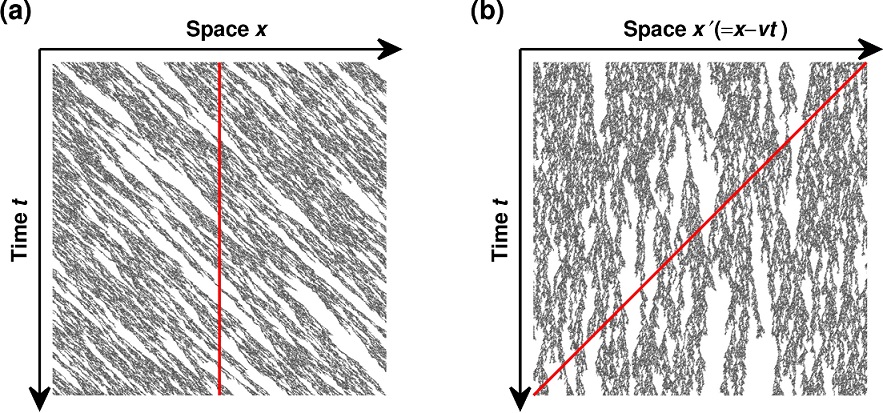}
        \caption{Typical spatiotemporal dynamics of DP with advection in the high-density regime. Active sites and inactive sites are depicted in gray and white, respectively. (a) In the original coordinate system, measuring inactive intervals at fixed position corresponds to measure inactive intervals along a vertical line (red line is shown as an example). (b) Same dynamics as (a), but Galilean transform is applied. The resulting picture is very similar to ordinary DP. In this coordinate system, the red line is tilted, whose angle depends on the velocity $v$ of the advection.}
        \label{advectschem}
    \end{figure}

    Simple heuristics provide a clear interpretation of the results, even though analytical treatment of the interval distribution is substantially more difficult than the order parameter. Typical spatiotemporal dynamics of DP with advection far away from the active wall is shown in figure\ \ref{advectschem}(a). If the distance from an active wall is sufficiently large, correlation between the local state at the probe and active wall is negligible, and as hence the dynamics look similar to each other if we apply suitable Galilean transform (figure\ \ref{advectschem}(b)), although active wall is still needed to sustain a high-density phase. Then, in this coordinate system, measuring the temporal intervals at the fixed position corresponds to measure the length of the inactive cluster along a tilted line. Here, we apply a somewhat rough approximation where we regard each cluster of inactive sites as an ellipse which has a major axis of length $l_{||}$ and a minor axis of length $l_{\perp}$. Choice of the major axis is based on the fact that $\nu_{||}>\nu_{\perp}$ in the DP universality class and so that the correlation time diverges more rapidly than the correlation length. Then, the length $L$ of the longest line across the ellipse gives the characteristic time for decay of the distribution of the inactive intervals. Hence a simple geometric argument gives
    \begin{equation}
    \label{oblique-longest}
    L\sim\sqrt{\frac{2\xi_{||}^2\xi_{\perp}^2}{v^2\xi_{||}^2+\xi_{\perp}^2}}.
    \end{equation}
    An important consequence of (\ref{oblique-longest}) is that, in the presence of the advection, $L$ is expected to exhibit power-law divergence with exponent $\nu_{\perp}$ near the critical point. The crossover is expected for $\varepsilon_*$ such that $v^2\xi_{||}^2\sim\xi_{\perp}^2$, that is,
    \begin{equation}
    \varepsilon_*\sim v_x^{\frac{1}{\nu_{||}-\nu_{\perp}}}.
    \end{equation}
    The parallel argument to what we gave for deriving (\ref{opcrit-scaling}) leads us to the following universal scaling ansatz for the characteristic time $\xi$:
    \begin{equation}
        \label{xiscaling}
        \xi(\varepsilon;\tan \theta)\sim \varepsilon^{-\nu_{||}}g((\tan\theta)^{-\frac{1}{\nu_{||}-\nu_{\perp}}}\varepsilon),
    \end{equation}
    where $g$ is the universal scaling function. We rescaled the data in figure \ref{decaylength}(b) according to this ansatz, and we again found a reasonably good collapse as a result, as shown in the inset of figure \ref{decaylength}(b). This suggests that the aforementioned simple argument captures the essence of the numerical results. Note that, if one would like to consider spatially $d$-dimensional systems, repeating the geometric argument in a case of ellipsoids instead of ellipses suffices, although the essential result is not expected to change (note that we have seen in Ref.\ \cite{Sano2016} that the essential results does not change in $d=2$ for $\theta=\pi/4$).    

\begin{figure}[tbp]
        \centering
        \includegraphics[width=\textwidth]{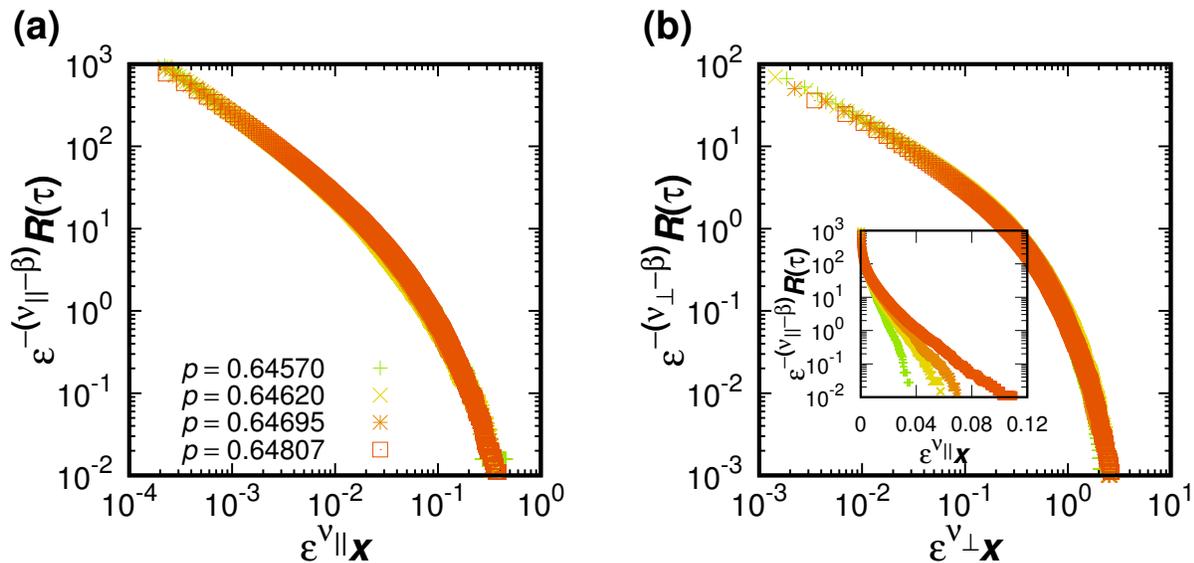}
        \caption{Universal scaling of the temporal interval distribution of the directed percolation model with active wall and asymmetric connection. (a) Temporal interval distributions for $\theta=0$, rescaled according to the scaling ansatz (\ref{interval-scalinghypo-CDF}). (b) Temporal interval distributions for $\theta=\pi/9$ (original data is shown in figure \ref{decaylength}a), rescaled according to the scaling ansatz (\ref{interval-scalinghypo_spatial-CDF}). Inset shows the same data rescaled instead according to (\ref{interval-scalinghypo-CDF}). The theoretical values of the critical exponents in the DP universality class were used to rescale the data.}
        \label{inact-univ}
    \end{figure}

    Recalling the phenomenological scaling argument given in section 2, it is natural to speculate that the universal scaling ansatz (\ref{cptempscaling}) may hold in a case without advection whereas (\ref{cpspatscaling}) with advection, as long as the system is sufficiently close to the critical point. To examine this speculation, we rescaled the data for $\theta=0$ and $\theta=\pi/9$ (as an example) according to the universal scaling ansatz (\ref{interval-scalinghypo-CDF}) and (\ref{interval-scalinghypo_spatial-CDF}), respectively. The results shown in figure\ \ref{inact-univ}(a) and \ref{inact-univ}(b) indicates this is indeed the case (It is important to note that, if the advection is present, the interval distribution for various $p$ does not overlap under rescaling with respect to the scaling ansatz (\ref{cptempscaling}), as shown in the inset of figure \ref{inact-univ}(b)). From these observations, one can safely conclude that the critical exponent $\nu_{\perp}$, in this case, can be measured through temporal interval distributions at the steady state, even if the advection is too strong for the crossover of $\rho(x)$ to be observed clearly. Hence we no longer need to rely on measuring the velocity of the activity wave \cite{Costa2010} which suffers from the same drawbacks as the critical spreading protocol.

    Since the irrelevance of advection (in an absence of a fixed boundary) is not peculiar to the DP universality class but is a feature shared among the universality classes which can be described in terms of single-species bosonic field theory \cite{Park2005}, we expect the methodology proposed in the present work to be applicable to some other universality classes of absorbing phase transitions (e.g. parity-conserving universality class), although one should be considerate of the relevance of advection in general (pair contact process with diffusion \cite{Henkel2004} is one counterexample). We leave the detailed investigation in this direction for future studies.

    \section{Concluding remarks}\label{discussion}
    Given the results we outlined in this work, now let us discuss an interpretation of the results of experimental literature concerning possible relation to the DP universality class. The experimental results prior to Takeuchi \textit{et al.}\ \cite{Takeuchi2007, Takeuchi2009} have two features in common: First, as is also pointed out by Henkel \textit{et al.}\ \cite{HenkelBook}, the largest deviation from the theoretical value of DP was found for estimate of the critical exponent $\nu_{||}$ associated with divergence of correlation time. Second, the estimated critical exponent $\nu_{||},\nu_{\perp}$ are smaller than the theoretical value. Our work suggests a likely explanation to this situation: If we measure the interval distribution using the empirical estimator (\ref{empirical-estimator}) under a short observation window, estimated correlation length/time saturates near the critical point (figure\ \ref{cptempscaling}(a)), which leads one to significant underestimation of the critical exponents. Recalling that effectiveness of the empirical estimator is a matter of comparison between the size of the window and the intrinsic correlation length/time of the system, it is reasonable that the estimation of $\nu_{||}$ is more strongly subject to the impact of a finite observation window than that of $\nu_{\perp}$ is. 
    
    To further substantiate our argument, let us revisit an earlier experimental work by Rupp \textit{et al.} \cite{Rupp2003} as an example. In their work, the authors estimated the correlation length (time) through the distribution of spatial (temporal) interval and the resulting critical exponent $\nu_{\perp}$ was consistent with theoretical value, whereas $\nu_{||}$ was massively smaller. However, at $\varepsilon\sim 3\times 10^{-2}$ or less, the estimated correlation time is comparable or longer than one-tenth of the observation window, and qualitatively consistent behavior can be seen if one discards that region as unreliable. Hence, it could be that the correlation time of the system indeed diverged in the same manner as predicted by the DP universality class but that the authors gave a biased estimate of the correlation time near the critical point due to a finite observation window. Thus, one should be cautious when performing and interpreting the measurements on a correlation time and a correlation length through the interval distributions. The estimator we employed in section \ref{ordinary-case} provides a practical solution to the issue.
    
    Considering the aforementioned delicacy of the problem of estimating the distributions, performing experiments with a system supplemented with active boundary condition and advection is advantageous, as we demonstrated in section \ref{numerics}: With this setup, one can sidestep the problem by measuring a correlation time through spatial decay of an order parameter, with a price of necessity to prepare the system so that it is sufficiently large at least in streamwise direction. Another important advantage is that an active wall prevents the system from accidentally falling into an absorbing state, making the situation simpler than ordinary systems in turn. Moreover, universal scaling ans\"{a}tze (\ref{opcrit-scaling}), (\ref{xiscaling}) provide another means to examine the universal features of absorbing phase transitions beyond a na\"{i}ve comparison between the estimated critical exponents and the theoretical value, if one can systematically adjust the strength of the advection. Thus we believe that experiments with an active wall and advection pave a way for quantitatively studying absorbing phase transitions in a laboratory.

    Feasibility of the procedure proposed in section \ref{numerics} should not be underestimated. As we discussed in section \ref{introduction}, advection is inevitably accompanied to spatiotemporal dynamics in some experimental systems. Also, an active wall can be reasonably implemented, provided that the control parameter of the system can be tuned locally. In a case of the turbulent liquid crystals, for example, one can achieve an active wall by placing a ``source'' region, where high voltage is applied, next to a region of interest. This can be done using separate electrodes for different regions. Similarly, the procedure is expected to be feasible in other experimental systems: In fact, our previous work \cite{Sano2016} was based on this idea, although extensive study about this situation as presented in this paper had not been performed at that time. We hope that the present work triggers further experimental effort of characterizing absorbing phase transitions, which, we believe, constitutes a significant step towards understanding of complex spatiotemporal dynamics in real physical systems in terms of non-equilibrium statistical mechanics.    
        
    To summarize, we proposed two ways to examine the universal features of absorbing phase transitions (the DP universality class in particular) properly and efficiently in experiments. Interesting directions for future studies, in our opinion, include (a) performing experiments in various systems (including those have been considered in literature) with proper statistical analysis to clarify how the DP universality class is robust in reality , (b) examining the universal scaling properties of DP with an active wall and advection by experiments, and (c) studying critical phenomena of the DP universality class with an active wall and advection on a more rigorous theoretical footing.
    
    \ack
    This work was supported in part by Grant-in-Aid for JSPS Research Fellow (Grant No.\ JP16J08207) and Grant-in-Aid for Scientific Research on Innovative Areas ``Fluctuation \& Structure'' (Grant No.\ JP25103004) from Japan Society for the Promotion of Science.
    
    \appendix
    \section{Remarks on a previous attempt to remove a bias in estimation of interval distributions}
\begin{figure}[tbp]
        \centering
        \includegraphics[width=0.5\textwidth]{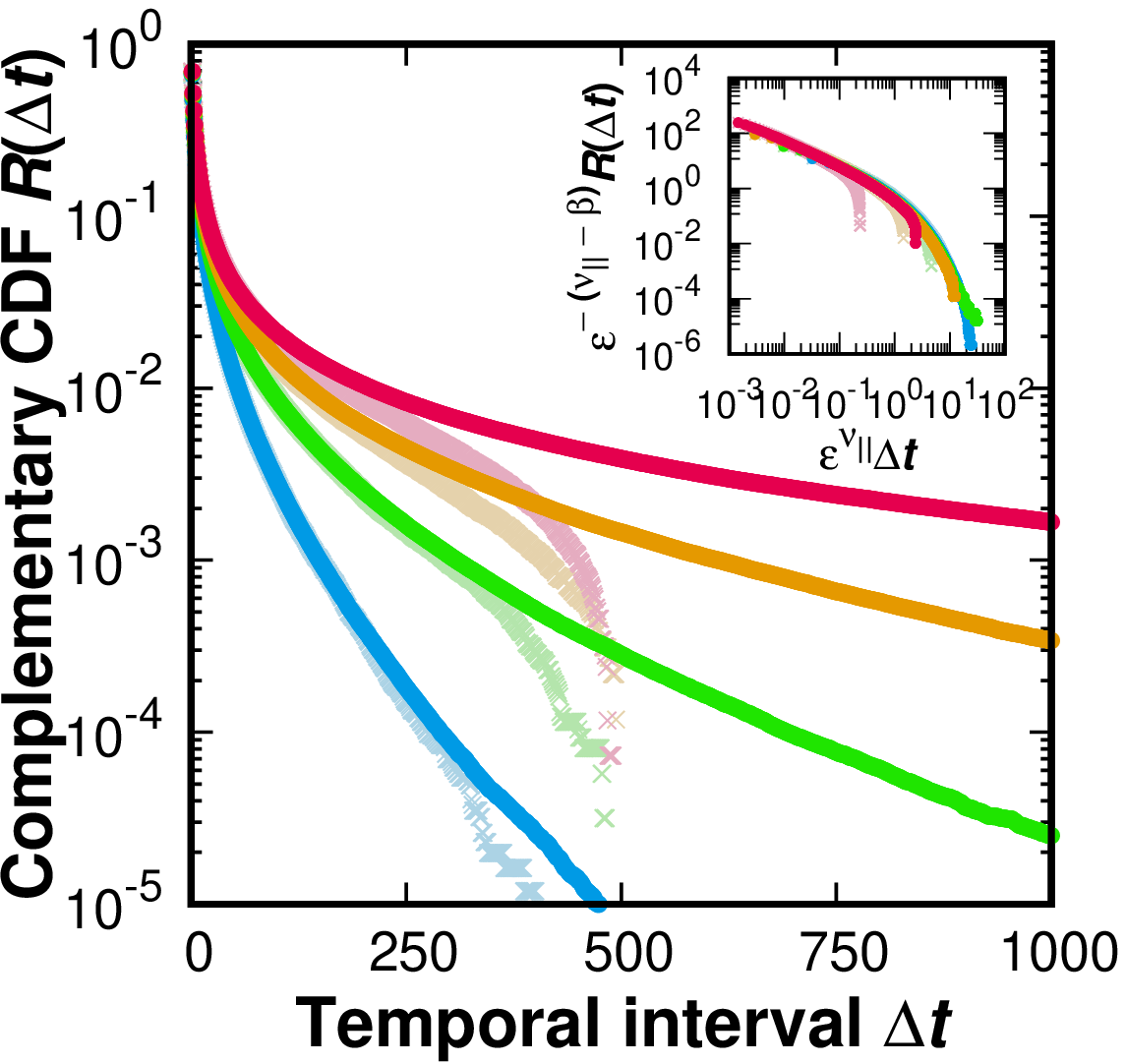}
        \caption{Temporal interval distribution of spatially one-dimensional contact process for $\lambda=3.33785$ (magenta), $\lambda=3.41098$ (orange), $\lambda=3.52412$ (green), and $\lambda=3.75039$ (blue). The distribution was estimated via the estimator proposed by Takeuchi \textit{et al.} ((\ref{corrected-numbers})--(\ref{corrected-CDF}) in the text). Results of the estimation under the window of $W=500$ is painted in light color whereas ones under the window of $W=5,000$ is painted in vivid color, and the inset shows the result of rescaling of the plot according to the scaling ansatz (\ref{interval-scalinghypo-CDF}). The theoretical values of the critical exponents in the DP universality class were used to achieve the data collapse.}
        \label{correctedscaling}
    \end{figure}

    This Appendix is devoted to examine an attempt to remove a bias caused by a finite observation in earlier experimental works, or more specifically, that made by Takeuchi \textit{et al.} \cite{Takeuchi2009} As described in the main text, Takeuchi \textit{et al.}\  divided the number $N(\Delta l)$ of observed spatial (temporal) intervals of size $\Delta l$ by $1-\Delta l/W$: Formally stating, they defined ``unbiased'' estimate $N_1(\Delta l)$ of the number of observed intervals by
    \begin{equation}
      \label{corrected-numbers}
      N_1(\Delta l)=\frac{N(\Delta l)}{1-l/W},
    \end{equation}
    where $N(\Delta l)$ and $W$ respectively denote the number of completely observed intervals of length $\Delta l$ and size of an observation window. To see whether this method is unbiased, we introduced a new estimator $\hat{P}_1(\Delta t;\varepsilon)$ for temporal interval distribution as
    \begin{equation}
      \label{corrected-distribution}
      \hat{P}_1(\Delta t;\varepsilon)=\frac{N_1(\Delta t;\varepsilon)}{\sum_{\Delta t} N_1(\Delta t;\varepsilon)},
    \end{equation}
    and computed the estimated complementary cumulative distribution function $\hat{R}_1(\Delta t;\varepsilon)$ by
    \begin{equation}
      \label{corrected-CDF}
      \hat{R}_1(\Delta t;\varepsilon)=\sum_{\Delta t^{\prime}=\Delta t}^{W}\hat{P}_1(\Delta t^{\prime};\varepsilon)
    \end{equation}
    in a case of the contact process. Apart from an estimator to use, we performed the analysis in the same way as described in the main text. The results shown in figure \ref{correctedscaling} indicate that the estimator used by Takeuchi \textit{et al.}\ is still biased, although the bias is much weaker than that caused by the empirical one (shown in figure \ref{cptempscaling}(a)).

\providecommand{\newblock}{}


\end{document}